\documentclass[aps,epsfig,twocolumn,showpacs]{revtex4}
\usepackage{amsmath}
\usepackage{graphicx}

\newcommand{\beq}{\begin{equation}}
\newcommand{\eeq}{\end{equation}}

\newcommand{\bea}{\begin{eqnarray}}
\newcommand{\eea}{\end{eqnarray}}

\begin{document}

%\title{Heat flux operator: Applications and conditions for current conversation}
\title{Heat flux operator, current conservation and the
formal Fourier's law}
% More "nano"
\author{Lian-Ao Wu and Dvira Segal}
\affiliation{Chemical Physics Theory Group, Department of Chemistry, and Center for
Quantum Information and Quantum Control, University of Toronto, \\
80 St. George Street, Toronto, Ontario M5S 3H6, Canada}

\begin{abstract}
By revisiting previous definitions of the heat current operator,
we show that one can define a heat current
operator that satisfies the continuity equation for a general
Hamiltonian in one dimension. This expression is useful for
studying electronic, phononic and photonic energy flow in linear
systems and in hybrid structures. The definition allows us to deduce
the necessary conditions that result in current
conservation for general-statistics systems. The discrete form of
the Fourier's Law of heat conduction naturally emerges in the present
definition.
\end{abstract}

\maketitle

\section{Introduction}

% % defintion is useful for studying heat transfer in nanosystems.
% fermionic, bosonic, Spin boson.
% finite systems. Contacts.

% Heat trasnfer in nanoscystems is interesting recently - oging into the quantum limit.
%There is a need for a basic model independent defintion, also not stationary.
%
% We discuss the correct defintion
% applies it for various systems

% background
The problem of heat transfer, electronic, phononic and photonic, in
molecules and nanosystems has recently gained lots of interest
\cite{Kim, Braun, Segalman, RectifE, Dlott, films,PekolaNat}. In
molecules, understanding heat flow is crucial for controlling
reactivity, molecular dynamics, and kinetics \cite{Uzer}. In
nanosystems, heat transfer has recently attracted much attention
with implications in thermal machinery
\cite{Tunable,Pekolaref,Pekolapump}, information processing and
computation \cite{Gate,Circuit}, and molecular-based
thermoelectricity \cite{Reddy,Hochbaum,Heath}. Of special interest
are hybrid structures, e.g. normal metal-superconductor junctions
with applications in thermometry and refrigeration \cite{Opport},
and atom-radiation field systems,  serving as a prototype for
studying thermodynamics of quantum systems \cite{Geva,Tannor}.

% Add example of spins
% Write more. Pekola. Tannor.
% heat flow in normal metal-superconductors junctions for applications?
% Tannor: Q thermodyanmics
% machine

% theory
From the theoretical point of view systems of interest include
collections of bosons, fermions, spins, and mixed-statistics models
\cite{ZotosRev}. For example, heat transfer from a dielectric solid
into a molecule may be studied using a spin-boson model where the
molecule is represented by a single anharmonic mode (spin) and the
bulk includes a collection of harmonic modes (boson)
\cite{SegalRectif}. In the analogous spin-fermion model an electronic
excitation is transferred between two metals through a local
mode, modeling a vibrating molecule.
If the central mode is harmonic, the model may further describe
radiative heat transfer between electronic conductors \cite{PekolaNat,radiation}.

% Our defintiion
In order to perform first principle quantum-mechanical calculations
of heat transfer in nanosystems it is necessary to consider a
model-independent non-perturbative definition of the heat current.
This expression should be applicable in non-stationary cases, as
well as in steady-state situations. While there is no unique
definition of the heat current operator in non-relativistic systems
\cite{Allen93}, the constructed expression should still fulfill a
symmetry requirement, as we discuss below. We present here a
consistent definition for the heat flux operator using a generic
one-dimensional (1D) Hamiltonian. We show that this expression is
useful for studying  vibrational, electronic and spin mediated heat
transfer, and that it yields a non-perturbative expression for the
heat current in hybrid systems, e.g. at a solid-molecule-solid
interface represented by a two-bath spin-boson model.

Furthermore, the definition also brings in a useful physical insight:
We derive a {\it necessary} condition
for energy conservation in various systems, bosonic and electronic,
by calculating the commutator
of the total flux operator with $H$, the total Hamiltonian.
If the current is a conserved quantity, the transport is ballistic,
the conductivity diverges, and Fourier's law of heat conduction cannot be
fulfilled \cite{Bonetto00}.

Derivation of the Fourier's law from fundamental principles,
classical \cite{Bonetto00,Lepri,Pereira06,Antti}, or quantum
\cite{Michel05,Michel06,WuSegal07}, is a great challenge in
theoretical physics. Model calculations manifested that the onset of
diffusional behavior delicately depends on the details of the
system. It is still not clear what necessary and sufficient
conditions must the Hamiltonian fulfill for showing the Fourier's
dynamics.
%
%While the law is obeyed by most macroscopic systems
%there exist families of microscopic Hamiltonians that allow heat transport
%without resistance. One of the members in the family is the system of Harmonic
%oscillators, in which the complete heat current is conserved, so that once
%prepared, a current in a closed loop system will never vanish.
% Our work on Fourier
%
Here we circumvent this challenge, and rather than test the
applicability of the Fourier's law in  specific systems, derive a
general, {\it necessary} condition for current conservation. Systems
that do not obey this condition may satisfy the Fourier's law. As an
example, we verify that in systems of harmonic oscillators the total
heat current is conserved, so that once prepared, a current in a
closed loop system will never vanish.
%This analysis thus sets the first step towards the exploration of the
%validity of Fourier's law of heat conduction in Hamiltonian systems.

% Thermal conductance
Another implication of the proper definition of the heat current is
the identification of a microscopic expression for the thermal
conductivity in terms of Hamiltonian parameters. This expression
might be useful for studying the thermal conduction properties of
molecular wires and spin chains.

%In this paper, we are interested in the definition of heat flux operator for
%a one-dimensional generic Hamiltonian, the Fourier's law deduced from our
%definition of the heat flux operator, the conditions for Hamiltonians that
%are able to provide conserved currents. Since there is no unique definition
%\cite{Allen93} for current, we shall first define a heat flux (current)
%operator and use it in all discussions.

The paper is organized as follows. In Section II we discuss the general
definition of the heat flux operator in one dimension. Section III applies
this expression to complex structures, e.g. the spin-boson model and
the spin-fermion model, prototype models for studying heat transfer
in hybrid systems. In Section IV we show that a current
conservation condition naturally emerges from the heat flux
definition for both bosonic and fermionic Hamiltonians. Section V
further explores current conservation in general 1D systems. From
the heat flux expression the discrete Fourier's law can be naturally
identified, as shown in Section VI. In Section VII we conclude.

%---------------------------------------------------------
\section{\protect\bigskip Definition of the energy flux operator
for a general Hamiltonian in one dimension}

Defining a heat flux operator for a specific system such as phonons
dates back to Hardy's early work \cite{Hardy63}. The idea was
applied to spin chains, see e.g.
\cite{Saito95,Saito96,Saito03,Casati05}, and other 1D systems, see
e.g. Ref. \cite{Michel06}. A general flux (current) operator may be
obtained by assuming that there exists an operator continuity
equation, for instance $\frac{\partial h(x,t)}{\partial
t}+\frac{\partial j(x,t)}{\partial x}=0$ in one dimension, where
$h(x,t)$ is the energy density operator and $j(x,t)$ is the heat
flux operator. For an $N$-site chain with $M$-states at each site,
one can introduce a workable definition of the energy density
operator, $h(x,t)=\sum_{s}h_{s}\delta (x-x_{s})$, where $h_{s}$ is a
discrete energy density operator of the $^{s}$th site. The total
Hamiltonian of the chain is therefore given by $H=\int
dxh(x,t)=\sum_{s}h_{s}$. Similarly, the heat flux can be written as
the a of localized contributions $j(x,t)=\sum_{s}j_{s}\delta
(x-x_{s})$, so that the continuity equation can be written in a
discrete form,
\begin{equation}
\frac{dh_{s}}{dt}=\frac{j_{s-1}-j_{s}}{a},
\label{eq:eq1}
\end{equation}
where $a$ is the lattice spacing and $j_{s}/a$ is the current
operator. The time evolution of $h_{s}$ in the Heisenberg
representation satisfies the Heisenberg equation of motion,
$\frac{d}{dt} h_{s}=i[H,h_{s}]$, assuming $h_s$ does not depend
on time in the Schr\"odinger representation ($\hbar\equiv 1$). This yields
\begin{equation}
\frac{dh_{s}}{dt}=i\sum_{k}[h_{k},h_{s}].
\label{eq:eq2}
\end{equation}
%
%Comparing this result with Eq. (\ref{eq:eq1}), we find that we can naturally define the
%current operator as
%%
%\begin{equation}
%j_s=\sum_{k>s} j_{s\rightarrow k};\,\,\,\,\,\,
%j_{s\rightarrow k}=ia[h_{s},h_{k}].
%%-\frac{ia}{2}[\Delta h_{s},e_{s}],
%\label{eq:eq3}
%\end{equation}
%
%where $\Delta h_{s}=h_{s+1}-h_{s}$ and $e_{s}=h_{s+1}+h_{s}$.
%This is a non-perturbative compact definition of the heat flux operator for the most
%general one-dimensional Hamiltonian.
%It satisfies the symmetry requirement $j_{s\rightarrow k}=-j_{k\rightarrow s}$.
In general, Eq. (\ref{eq:eq2}) cannot be expressed in terms of the
difference of two operators at two sites as in (\ref{eq:eq1}), yet
we can identify the currents $j_s$ and $j_{s-1}$ for a specific
model Hamiltonian. We use here a generic 1D Hamiltonian with up to
two-body nearest-neighbor interactions,
\bea
H=\sum_{s=1}^{N}(h_{s}^{0}+V(s,s+1)),
\label{eq:H}
\eea
where $h_{s}^{0}$ is the local Hamiltonian at site $s$.
While the local energy density can not be uniquely defined \cite{Allen93},
one could make a reasonable separation of $V$, and assign  mixed terms half to
site $s$ and half to $s+1$. With this partition
the energy density at the ${s}$th local site becomes
\begin{equation}
h_{s}=h_{s}^{0}+\frac{1}{2}\left[ V(s,s+1)+V(s-1,s)\right].
\label{eq:eq4}
\end{equation}
This equation satisfies $H=\sum_{s=1}^{N}h_{s}$, as
required, when one sets $V(N,N+1)=V(0,1)=0$.
For this Hamiltonian, the heat flux operator can be identified as
\begin{equation}
j_{s}=j_{s\rightarrow s+1}^{(2)}+j_{s}^{(4)},
\label{eq:js}
\end{equation}
where
\begin{equation}
j_{s\rightarrow s+1}^{(2)}=\frac{ia}{2}[(h_{s}^{0}-h_{s+1}^{0},V(s,s+1)],
\label{eq:j2}
\end{equation}
is a two-site contribution and
\bea
j_{s}^{(4)} &=&\frac{ia}{2}\big \{[V(s,s+1),V(s+1,s+2)] \nonumber\\
&&+[V(s-1,s),V(s,s+1)]\big \}
\label{eq:j4}
\eea
is an operator connecting four sites, accounting for higher order
inter-site interaction terms. As we show below, in some cases it is
exactly zero. It is also noticeable that in our case Eq.
(\ref{eq:eq2}) could be written in terms of the difference of
operators at two neighbor sites. The definition also naturally
classifies the perturbative orders with respect to the inter-site
coupling $V$: The order of the flux operator (\ref{eq:j4}) is higher
than that of (\ref{eq:j2}).

The definitions (\ref{eq:js})-(\ref{eq:j4}) possess significant
symmetric features. For instance, $j_{s\rightarrow s+1}^{(2)}$
trivially shows the exchange symmetry $j_{s\rightarrow
s+1}^{(2)}=-j_{s+1\rightarrow s}^{(2)}$, assuming
$V(s,s+1)=V(s+1,s)$. The exchange symmetry is an essential
requirement when defining a current operator, since the current in
opposite directions must have the same absolute value. The operators
$j_{s}^{(4)}$ has a similar exchange property, but four sites are
involved.

The definition (\ref{eq:js}) is state- and symmetry-independent
unlike the expression utilized in Refs.
\cite{Michel06,Gemmer06,Gemmer07,Michel08},
$j_s=ia[h_s^0,V(s,s+1)]$,
 which requires that the
Hamiltonian fulfills the symmetric condition
$[h_{s}^{0}+h_{s+1}^{0},V(s,s+1)]=0$, see Appendix A for details. In
order to increase generality, Ref. \cite{Gemmer06} further suggests
a 'symmetrized local flux' that has the same form as
$j^{(2)}_{s\rightarrow s+1}$.

The heat flux operator was also extensively examined in 1D chains {\it in
the absence of an on-site energy term } ($h_s^0=0$), e.g. the
Heisenberg model at zero magnetic field \cite{Zotosoperator}. In
this case the energy at each site was defined as $h_s=V(s,s+1)$,
leading to the current operator $j_{s-1}=ia[V(s-1,s),V(s,s+1)]$.
Since in this paper we are interested in the opposite limit, i.e. in
structures where the inter-site interaction is considered as a
perturbation to the local energy, e.g. impurity models, the choice
(\ref{eq:eq4}) for the local energy is more appropriate.

Note, that we could have also defined a high order local interaction
term $U(s)$. For phononic systems $U$ includes on-site interactions,
incorporating harmonic and anharmonic contributions. For fermionic
systems $U$ may represent a local electron-electron repulsion. The potentials
$V(s,s+1)$ and $V(s,s+1)+U(s)+U(s+1)$ indeed produce different flux
operators. We adopt here the convention that local $s$ interactions
(one-body and many-body) are all included within the potential
$h_s^0$.

Finally, one could consider next nearest neighbor interactions, and
by following the same procedure, identify the current $j_s$.
%However, the additional
%term, $\frac{ia}{2}[(h_{s}^{0},U(s)]$, is a local physical quantity
%and does not provide a significant contribution to
%the current (XXX). Therefore, one should exclude such terms in the potential $%
%V(s,s+1)$ in defining a current operator (XXX).

%-----------------------------------------
\section{Current operator in hybrid structures}

The definition (\ref{eq:js}) can be applied to non-identical
interacting systems which are spaciously connected. For example, we
may consider an impurity spin coupled to two solids, and study the
heat current at the contact. The bulk, serving as a thermal
reservoir, may be composed of electrons (the Kondo problem)
\cite{Kondo}, collections of harmonic modes (the spin-boson model)
\cite{Legget}, or spins \cite{Stamp,Fazio}. This impurity-bath
scenario is the standard in molecular electronics and nanomechanical
experiments, where the heat transfer properties of a molecule
connected to solid or liquid interfaces are investigated
\cite{Braun,Segalman,RectifE,Dlott}.

The generic impurity-bath Hamiltonian includes a central unit $H_{spin}$,
two independent reservoirs $H_{\nu}^0$ ($\nu=L,R$) maintained at
different temperatures, and system-bath couplings $V_{\nu}$. The
heat flux operator, e.g. at the $L$ contact is given by Eqs.
(\ref{eq:js})-(\ref{eq:j4}), disregarding for convenience the
lattice constant $a$. Assuming that $[V_L,V_{R}]=0$ we find that the
current from the $L$ contact into the junction is given by
\bea
j_{L}=\frac{i}{2}[H_L^0-H_{spin},V_L].
\label{eq:jimpurity}
\eea
We apply next this result assuming either bosonic baths or electronic reservoirs.

\textit{Spin-Boson model.--- }
%As an example we first consider a spin-boson model describing a two level
%system coupled to a bath of harmonic oscillators.
A two-level system connected to two harmonic baths held at different
temperatures serves as a prototype model for investigating phononic
transfer in a nonlinear molecular junction. Calculations at the
level of the Master equation, assuming  weak system-bath couplings
while ignoring coherence effects, have revealed interesting
dynamics, e.g. thermal rectification \cite{SegalRectif}, negative
differential resistance \cite{SegalNDR}, and pumping of heat
\cite{SegalPump}. It is of interest to derive a general expression
for the heat current which is not limited to the weak coupling
limit. Such an expression will open the door for non-perturbative
calculations of heat current in strongly coupled molecular systems.
The multi-bath spin-boson Hamiltonian is given by
\bea
H_{SB}= \frac{B}{2}\sigma^z+ \sum_{\nu,q}\omega_q
b_{\nu,q}^{\dagger} b_{\nu,q} +\sigma^x \sum_{\nu,q}
\lambda_{\nu,q}(b_{\nu,q}^{\dagger}+b_{\nu,q}).
\nonumber\\
\eea
Here $\sigma^{i}$ ($i=x,y,z$) are the Pauli matrices and $B$ is the spin splitting.
The reservoirs ($\nu=L,R$) include two infinite sets of harmonic oscillators (creation operators $b_{\nu,q}^{\dagger}$).
Spin-bath interaction strength is denoted by $\lambda_{\nu,q}$, possibly different at the two ends.

Let $H_{\nu}^0$ denotes the local Hamiltonian of the ensemble of
harmonic oscillators at the $\nu$ boundary, $H_{spin}=
\frac{B}{2}\sigma^{z}$ be the Hamiltonian of the spin and
$V_{\nu}=\sum_q \lambda_{\nu,q}
\sigma^{x}(b^{\dagger}_{\nu,q}+b_{\nu,q})$ be the interaction. Using
Eq. (\ref{eq:jimpurity}), the energy flux from the $L$ contact to
the spin unit is given by
\bea
j_{L}&=&\frac{1}{2}
\big[ i\sigma^x\sum_{q}\omega_q\lambda_{L,q}(b_{L,q}^{\dagger}-b_{L,q})
\nonumber\\
&+&
B \sigma^{y}\sum_q\lambda_{L,q}(b_{L,q}^{\dagger}+b_{L,q})
\big],
\label{eq:JSB}
\eea
or equivalently,
\bea
j_{L}=\frac{1}{2}[B\sigma^y X_L + \sigma^x P_L],
\label{eq:JSB2}
\eea
where $X_L=\sum_{q}\lambda_{L,q}(b_{L,q}^{\dagger}+b_{L,q})$ and
$P_L=i\sum_{q}\lambda_{L,q}\omega_q(b_{L,q}^{\dagger}-b_{L,q})$.
An analogous expression exists at the $R$ side.
It can  be shown that the flux
operator (\ref{eq:JSB}) reduces to the stationary heat flux expression
utilized in Refs. \cite{SegalRectif, SegalNDR, SegalPump} when
system-bath couplings are weak and the Markovian limit is assumed,
\bea
\langle j_{L}\rangle=-B [ k_{u\rightarrow d}^Lp_{u}-k_{d\rightarrow u}^Lp_{d} ].
\label{eq:jmark}
\eea
Here $\langle j\rangle$ denotes the trace over system and bath degrees of freedom,
 $p_{u}$ ($p_d$) is the steady state population of the up (down) spin
level and $T_{\nu}$ is the temperature at the $\nu$ contact. The
rate constants satisfy the detailed balance relation,
$k_{d\rightarrow u}^{\nu}=k_{u\rightarrow d}^{\nu}e^{-B/T_{\nu}}$,
where
\bea
k_{u\rightarrow d}^{\nu}&=&\int_{-\infty} ^{\infty}e^{iB\tau}\langle
X_{\nu}(\tau)X_{\nu}(0)\rangle d\tau
\nonumber\\
&=& 2\pi \sum_q
\lambda_{\nu,q}^2[n_B^{\nu}(\omega_q)+1]\delta(B-\omega_q).
\label{eq:rateb} \eea
$n_B^{\nu}(\omega_q)=\left[ e^{\omega_q/T_{\nu}}-1\right]^{-1}$ is
the Bose-Einstein distribution function with the Boltzmann constant
$k_B\equiv 1$. Equation (\ref{eq:jmark}) describes energy current at
the $L$ contact as the balance between an energy extraction from the
$L$ reservoir into the spin, and an energy loss from the spin to the
bath. Appendix B presents in details the derivation of this
perturbative result from the general operator
expression (\ref{eq:JSB2}).

Similarly, one may analyze the transport properties of the
diagonally coupled spin-boson model with
$V_{\nu}=\sigma^z\sum_q\kappa_{\nu,q}(b_{\nu,q}^{\dagger}+b_{\nu,q})$
and $H_{spin}=\frac{B}{2} \sigma^z +\frac{\Delta}{2} \sigma^x$,
leading to complicated behavior due to the non-separability of the
two reservoirs \cite{SegalRectif,SB}.

\textit{Spin-Fermion model.--- }
The spin-fermion model, where a spin impurity is coupled to two Fermi seas
of different temperatures and/or chemical potentials,
is another example of a hybrid structure, useful e.g. for studying electronic and radiative
heat transfer between metals \cite{radiation},
\bea
H_{SF}=\frac{B}{2}\sigma^z+ \sum_{\nu,k}\epsilon_k c_{\nu,k}^{\dagger}c_{\nu,k}
+\sigma ^x \sum_{\nu,k,q} \alpha_{\nu,k, q}c_{\nu,k}^{\dagger}c_{\nu,q}.
\label{eq:SF0}
\eea
The first term here accounts for spin splitting. The second term includes the two independent
reservoirs (leads) of spinless electrons, creation operator
$c_{\nu,k}^{\dagger}$, ($\nu=L,R$). We assume that the leads are kept (each) in thermal equilibrium
at temperature $T_{\nu}$ and chemical potential $\mu_{\nu}$.
The last term in (\ref{eq:SF0}) describes spin-bath interactions,
where we disregard charge tunneling between the metals and allow
only for transfer of energy excitations.
Utilizing Eq. (\ref{eq:jimpurity}), the heat current at the $L$ contact is given by
\bea
j_{L}&=&
\frac{i}{2} \Big[ \sigma^x \sum_{k,q}\epsilon_k \alpha_{L,k,q}(c_{L,k}^{\dagger}c_{L,q} -
c_{L,q}^{\dagger}c_{L,k} )
\nonumber\\
&-&iB \sigma ^y \sum_{k,q} \alpha_{L,k,q} c_{L,k}^{\dagger}c_{L,q}   \Big].
\label{eq:SF}
\eea
If the metals have strictly linear dispersion relation
this result can be exactly mapped into a bosonized description \cite{Luttinger}
to yield the current (\ref{eq:JSB}).
Deviations are expected when the metals have energy dependent density of states \cite{radiation}.
Following the derivation sketched in Appendix B, taking into account the fermionic nature
of the operators, one can show that
in second order system-bath coupling, going into the Markovian limit, the stationary heat current is
given by Eq. (\ref{eq:jmark}) with the rates
\bea
&&k_{d\rightarrow u}^{\nu}= \int_{-\infty}^{\infty}
e^{-iB\tau}\langle F_{\nu}(\tau) F_{\nu}(0)\rangle d\tau
\nonumber\\
&&=
2 \pi \sum_{k,q} |\alpha_{\nu,k,q}|^2n_F^{\nu}(\epsilon_k)
[1-n^{\nu}_F(\epsilon_q)]\delta(\epsilon_k-\epsilon_q-B)
\nonumber\\
&&k_{d\rightarrow u}^{\nu}=k_{u\rightarrow d}^{\nu}e^{-B/T_{\nu}},
\label{eq:ratef}
\eea
where
$F_{\nu}=\sum_{k,q}\alpha_{\nu,k,q}c_{\nu,k}^{\dagger}c_{\nu,q}$, is
the force the bath exerts on the system, and
$n_F^{\nu}(\epsilon)=[e^{(\epsilon-\mu_{\nu})/T_{\nu}}+1]^{-1}$ is
the Fermi-Dirac distribution function of the $\nu$ bath.

The perturbative rate expression (\ref{eq:jmark}) also holds for
mixed boson-fermion systems, e.g. when  energy is directed from a
phonon bath into an electronic excitation through a local impurity.
One simply employs then the expressions (\ref{eq:rateb}) and
(\ref{eq:ratef}) for the phononic and electronic bath-induced
transitions.

%--------------------------------------------------
\section{Current conservation conditions for bosonic and fermionic systems}

With the help of the heat flux operator we can obtain general properties
of specific quantum systems \cite{Zotosoperator}.
This is in contrast to standard calculations
where one needs to make use of specific quantum states \cite{Saito96,Saito03,Casati05}.
We prove next that linear harmonic systems and some special
spin chains ($XY$, Ising) have zero thermal resistance using the operator form of the energy flux.
%Here we first consider the system with one particle at each site of a chain.
%If the interaction between sites
%is linear, the flux operator can be expressed in terms of bosonic operators.
%(XXX)

\textit{Bosons.--- } We consider the quantized system used in
\cite{Lepri}, $h_{s}^{0}=p_{s}^{2}/2+U(x_{s})$ and inter-site
potential $V(s,s+1)=V(x_{s},x_{s+1})$, where $x_{s}$ and $p_{s}$ are
the coordinate and momentum of the particle at the ${s}$ site. It is
easy to show that $j_{s}^{(4)}=0$, thus the flux operator is given
by
\bea
j_{s}=\frac{a}{4}\left[ \left\{
{p_{s}, \frac{\partial V(s,s+1)}{\partial x_s}} \right\}_{+}
-
\left\{
{p_{s+1}, \frac{\partial V(s,s+1)}{\partial x_{s+1}}}
\right\}_{+} \right]
\nonumber\\
\eea
where $\left\{ {}\right\} _{+}$
denotes the anticommutation relation. This is just the quantized form of
the classical flux defined in reference \cite{Lepri}. For the the quadratic
interaction $(x_{s}-x_{s+1})^{2}$  we should
exclude the local terms $x_{s}^{2}$ and $x_{s+1}^2$,
or shift them into $h_s^0$ and $h_{s+1}^0$ respectively, as discussed in  Section II.
For a bilinear coupling model we thus consider the interaction
$V(s,s+1)=\lambda x_{s}x_{s+1}$ with spring constant $\lambda$. The flux operator then reads
\bea
j_{s} &=&a\lambda (p_{s}x_{s+1}-x_{s}p_{s+1})/2
\nonumber\\
&=&-ia\lambda (b_{s}^{\dag }b_{s+1}-b_{s+1}^{\dag }b_{s})/2,
\eea
where the second line is the bosonic expression with the creation (annihilation) operator
$b_{s}^{\dag}$ ($b_{s}$). The commutation relation
between the total Hamiltonian and the current operator is given by
\begin{equation}
\lbrack j_{s},H]=i\frac{a\lambda}{2}\left(
x_{s}\frac{\partial U(x_{s+1})}{\partial x_{s+1}}%
-x_{s+1}\frac{\partial U(x_{s})}{\partial x_{s}}\right)+\widehat{O}(\lambda ^{2}).
\label{eq:bosonJH}
\end{equation}
Therefore, if $\frac{\partial U(x)}{\partial x}=0$, or $U\propto x^2$,
$[j_{s},H]=0$ within first order coupling.
This implies that
free particle motion and harmonic potentials pertain a constant current, or
in other words, the heat current is conserved in these systems.

One can also calculate the higher order term in  (\ref{eq:bosonJH}),
$\widehat{O}(\lambda ^{2})=ia\lambda
^{2}(x_{s}^{2}+x_{s}x_{s+2}-x_{s-1}x_{s+1}-x_{s+1}^{2})/2$. If the total
flux is defined as $J=\sum_{s=1}^{N}j_{s}$,
the commutation relation between the complete flux and the total
Hamiltonian is given by $[J,H]=i\lambda ^{2}\frac{a}{2}(x_{1}^{2}-x_{N}^{2})$
for the harmonic $U\propto x^2$ potential.
Therefore, the Heisenberg equation of motion reads
\bea
dJ/dt=\frac{a\lambda ^{2}}{2}(x_{1}^{2}-x_{N}^{2}).
\label{eq:BCons}
\eea
This result shows that the total flux depends only on the contacts properties:
coupling strength and temperature (going into thermal averages).
Furthermore, in closed loop systems, the complete
current $J$ is a constant operator. This conclusion is well established, however,
we give here a simple proof of the operator form, without the need to
go into the system's quantum states.
It can be shown that the current is also conserved for {\it disordered} 1D harmonic systems.
For example, assuming different force constants between sites $\lambda_{s,s+1}$,
one gets $dJ/dt= \frac{a}{2}(\lambda_{1,2}^2 x_1^2-\lambda_{N-1,N}^2x_N^2)$.

\textit{Fermions: Nearest Neighbor Spin systems. } We consider next
a periodic spin chain of length $N$. The system can be mapped into a system of fermions
using the  Wigner-Jordan transformation, see e.g. \cite{WuLidar02}.
Let the on-site potential $h_s^0$ and the inter-site potential $V$ be
\bea &&h_{s}^{0}=\frac{\epsilon }{2}\sigma _{s}^{z}
\nonumber\\
&&V(s,s+1)=\lambda (A\sigma _{s}^{x}\sigma _{s+1}^{x}+B\sigma _{s}^{y}\sigma
_{s+1}^{y}+C\sigma _{s}^{z}\sigma _{s+1}^{z}),
\nonumber\\
\label{eq:spinH}
\eea
where $A,B,C$ are the interaction coefficients.
It is easy to show that the first-order flux operator is given by
\bea j_{s\rightarrow s+1}^{(2)}=a\epsilon \lambda
\frac{A+B}{2}(\sigma _{s+1}^{y}\sigma _{s}^{x}-\sigma _{s}^{y}\sigma
_{s+1}^{x}). \label{eq:spinJ} \eea
Using the Wigner-Jordan transformation, the current can be also rewritten as
\bea j_{s\rightarrow s+1}^{(2)}=ia\epsilon \lambda (A+B)(c_{s+1}^{\dag
}c_{s}-c_{s}^{\dag }c_{s+1}),
\label{eq:J2WJ}
\eea
expressed in terms of spinless fermionic creation and annihilation
operators $c_{s}^{\dag }$ and $c_{s}$ respectively. The second order
contribution $j_{s}^{(4)} \propto \lambda^2$ is nonzero in general,
but is too cumbersome to be included here.

The current operator $j_{s\rightarrow s+1}^{(2)}$ is essentially the
standard spin current operator multiplied by the bias $\epsilon$.
This term reflects energy flow due to spin current, while
$j_{s}^{(4)}$ accounts for thermal energy flow \cite{Zotosoperator}.
At weak inter-site coupling, $\lambda\ll \epsilon$, $j_{s\rightarrow
s+1}^{(2)}$  dominates the energy current, while for zero magnetic
fields only $j_{s}^{(4)}$ survives. Throughout the paper we always
assume nonzero magnetic splitting $\epsilon$, unless otherwise
stated.

We continue and analyze current conservation in the model
(\ref{eq:spinH}),
\bea &&[j_s,H] =[j_{s\rightarrow s+1}^{(2)},h_s^0+h_{s+1}^{0}]
+[j_s^{(4)},h_s^0+h_{s+1}^0] \nonumber\\ &&+ [j_{s\rightarrow
s+1}^{(2)},V(s-1,s)+V(s,s+1)+V(s+1,s+2)]. \nonumber\\ \eea
To the first order in $\lambda$ the commutator is therefore given by
\bea &&[j_{s\rightarrow s+1}^{(2)},h_s^0+h_{s+1}^0] \nonumber\\
&&=a\epsilon^2\lambda \frac{A+B}{4}[\sigma_{s+1}^y\sigma_s^x
-\sigma_s^y\sigma_{s+1}^x,\sigma_s^z+\sigma_{s+1}^z]=0.
\nonumber\\
\eea
Thus, for the periodic spin chains considered here, only high order
terms in $\lambda$ may lead to current decay.
We discuss next some special cases: (i) $A=-B$, corresponding to the
antiferromagnetic phase. Here $j_{s \rightarrow s+1}^{(2)}=0$,
implying that there is no current in the antiferromagnetic phase in
the first order approximation. (ii) The Heisenberg model, $A=B=C$.
In this case the flux operator (\ref{eq:j2}) agrees with the
definition of Ref. \cite{Michel06}, see also Appendix A, since
$[(\sigma _{s}^{z}+\sigma _{s+1}^{z},V(s,s+1)]=0$. This system was
extensively investigated in Refs. \cite{Michel06,Michel08}.
%(write some results?)
(iii) The \textit{XY} model, $A=B$ and $C=0$. We
calculate here the high order contribution to the current and find
\bea j^{(4)}_{s} &=&i\frac{a\lambda^{2}A^{2}}{2}(c_{s}^{\dag
}c_{s+2}-c_{s+2}^{\dag }c_{s} \nonumber\\ &&+c_{s-1}^{\dag
}c_{s+1}-c_{s+1}^{\dag }c_{s-1}),
\label{eq:J4XY}
\eea
Combining Eq. (\ref{eq:J2WJ})   with Eq. (\ref{eq:J4XY}) we get that
$[j_{s},H]=0+\widehat{O}(\lambda ^{2})$ in the $XY$ model.
The current operator is therefore a constant in the first order approximation,
while the total current exactly becomes
\bea
dJ/dt=8a\epsilon \lambda ^{2}A^{2}(n_{1}-n_{N}),
\label{eq:FCons}
\eea
in analogy with Eq. (\ref{eq:BCons}) for the bosonic Hamiltonian.
Here $n_{s}=c_{s}^{\dag }c_{s}$ is the number operator.
We conclude that the total current across the systems depends
only on the properties of the chain's ends. Thus, in
closed loop systems the total current $J$ is a constant operator.

As a final (iv) case we consider the transverse Ising model,
$B=C=0$. Here $j_{s\rightarrow s+1}^{(2)}=\frac{\lambda\epsilon
A}{2}(\sigma_s^x\sigma_{s+1}^y- \sigma_s^y \sigma_{s+1}^x)$,
$j_{s}^{(4)}=0$. The commutator $[j_{s},H]$ is zero in first order
of $\lambda$ while the second order contribution, resulting from the
commutator $[j_{s\rightarrow
s+1}^{(2)},V(s-1,s)+V(s,s+1)+V(s+1,s+2)]$, leads to
\bea
dJ/dt=4a\epsilon \lambda ^{2}A^{2}(n_{1}-n_{N}).
\eea

We can summarize our observations as follows:
If a Hamiltonian is written by a linear combination of bilinear
operators, a bosonic set $\left\{ b_{s}^{\dag }b_{t},b_{s}^{\dag
}b_{t}^{\dagger },b_{s}b_{t}\right\} $ and a fermionic set $\left\{
c_{s}^{\dag }c_{t},c_{s}^{\dag }c_{t}^{\dagger },c_{s}c_{t}\right\} $, it
can always be expressed in terms of quasiparticle operators
$\gamma _{q}$, where $H=\sum_{q=1}^{N}\epsilon _{q}\gamma _{q}^{\dag }\gamma
_{q}$ (see e.g. \cite{Ring}).
Since there is no interaction between the quasiparticles, the systems behaves like a
collection of free particles.
The harmonic oscillator chain with  linear couplings
is an example of bosonic Hamiltonian. The \textit{XY}
models are examples for independent fermions.
Both systems yield ballistic motion with no thermal resistance.
In contrast, the Heisenberg model with nonzero magnetic field does not belong to such systems because
it contains an on-site interaction
$c_{s}^{\dag }c_{s}c_{t}^{\dag}c_{t}$ when $C$ is not zero \cite{ZotosRev}.

%--------------------------------------------------------
\section{Necessary condition for current conservation for a general one-dimensional system}

We Consider a chain of length $N$ with $M$ levels at each site. The
commutation relation between the total Hamiltonian and the flux operator can be
written as
\bea
\lbrack j_{s},H]=F(\lambda )+\widehat{O}(\lambda ^{2}),
\label{eq:Flambda}
\eea
where $F(\lambda )$ is the first order term and $\widehat{O}(\lambda
^{2})$ contains higher order terms in $\lambda$.
The necessary condition for current conservation is $F(\lambda )\equiv 0$.
We emphasize that this is only a {\it necessary} condition. If $F(\lambda)\neq0$
the system potentially shows a diffusive dynamics.

The most general Hamiltonian for this system can be generated by
$M^{2}$ generator set
$g=\{\overrightarrow{h},E^{\overrightarrow{\alpha }}\}$,
 $\overrightarrow{h}=(n^{1},n^{2},...,n^{M})$ is the vector operator with
$n^{i}=\left| i\right\rangle \left\langle i\right|$.
$E^{\overrightarrow{\alpha }}$  denotes $M^{2}-M$ operators
$\left| i\right\rangle \left\langle j\right| $ where $i\neq j$. The
vectors $\overrightarrow{\alpha }$'s are $M$-dimensional,  and are
usually referred to as roots \cite{Wybourne}.
The commutation relation between $\overrightarrow{h}$ and
$E^{\overrightarrow{\alpha }}$ is
$[\overrightarrow{h},E^{\overrightarrow{\alpha
}}]=\overrightarrow{\alpha }E^{\overrightarrow{\alpha }}$, where the
vector $\overrightarrow{\alpha }$ can be considered as an eigenvalue
of the vector operator $\overrightarrow{h}$. For instance, in the
two level system ($M=2$), $\overrightarrow{h}=(\left| 1\right\rangle
\left\langle 1\right| ,\left| 2\right\rangle \left\langle 2\right|
)$, $E^{\overrightarrow{\alpha }}=(\left|1\right\rangle\left\langle
2\right|,  \left| 2\right\rangle \left\langle 1\right|  )$ and
there are two roots $\overrightarrow{\alpha }_{1}=(1,-1)$ and $%
\overrightarrow{\alpha }_{2}=(-1,1),$ corresponding to  $E^{\overrightarrow{%
\alpha }_{1}}=\left| 1\right\rangle \left\langle 2\right| $ and $E^{%
\overrightarrow{\alpha }_{2}}=\left| 2\right\rangle \left\langle
1\right|$. Appendix C presents the $M=3$ case.

Using this notation, the most general Hamiltonian up to a two-body
interaction can be written as
\bea
H=\sum_s \overrightarrow{\epsilon }\cdot \overrightarrow{h_s}+\lambda \sum_s [V_{%
\overrightarrow{\alpha },\overrightarrow{\beta }}E_{s}^{\overrightarrow{%
\alpha }}E_{s+1}^{\overrightarrow{\beta }}+V_{d}(s,s+1)]
\nonumber\\
\label{eq:Hg} \eea
where the vector $\overrightarrow{\epsilon }=(\epsilon _{1},\epsilon
_{2},...,\epsilon _{M})$,  $\epsilon _{i}$ is the ${i}$ state energy
level and $V_{\overrightarrow{\alpha },\overrightarrow{\beta }}$ are
inter-site coupling parameters. The units are assumed to have
identical spectra and we use constant nearest-neighbor interactions
along the chain. The last term in (\ref{eq:Hg}) includes many body
interactions $V_{d}(s,s+1)=\sum U_{i,j}n_{s}^{i}n_{s+1}^{j}$. It is
easy to show that the commutator of the current with $H$ yields
\bea
F(\lambda )=-\frac{ia\lambda }{2}\sum V_{\overrightarrow{\alpha },%
\overrightarrow{\beta }}[(\overrightarrow{\epsilon }\cdot \overrightarrow{%
\alpha })^{2}-(\overrightarrow{\epsilon }\cdot \overrightarrow{\beta }%
)^{2}]E_{s}^{\overrightarrow{\alpha }}E_{s+1}^{\overrightarrow{\beta }}.
\nonumber\\
\eea
The necessary condition for current conservation, $F(\lambda )\equiv 0$, therefore implies
\begin{equation}
\overrightarrow{\epsilon }\cdot \overrightarrow{\alpha }=\pm \overrightarrow{%
\epsilon }\cdot \overrightarrow{\beta }   \label{eq:condab}
\end{equation}
for nonzero coupling parameters $V_{\overrightarrow{\alpha
},\overrightarrow{\beta }}$.
 This condition (with the plus sign) is naturally fulfilled for harmonic systems, since
$(\epsilon_{j-1}-\epsilon_{j})=  (\epsilon_{k-1}-\epsilon_{k})$ for any $j,k$. For
fermionic models $M$=2 and the $\overrightarrow{\epsilon} \cdot
\overrightarrow{\beta}=-\overrightarrow{\epsilon}
\cdot\overrightarrow{\alpha}$ condition is trivially conformed.
Both systems indeed lead to current conservation, see Section IV.

For a system with an arbitrary spectra this condition translates
into $\overrightarrow{\alpha }=\pm \overrightarrow{\beta }$, implying that the interaction contains only the
following terms: $E_{s}^{\overrightarrow{\alpha
}}E_{s+1}^{\overrightarrow{\alpha }}$, $E_{s}^{
\overrightarrow{\alpha }}E_{s+1}^{-\overrightarrow{\alpha }}$ and
$V_{d}(s,s+1)$ \cite{noteE}. The corresponding current operator is
\bea j_{s \rightarrow s+1}^{(2)}=ia\sum V_{ \overrightarrow{\alpha
},-\overrightarrow{\alpha }}(\overrightarrow{\epsilon } \cdot
\overrightarrow{\alpha } ) E_{s}^{\overrightarrow{\alpha
}}E_{s+1}^{- \overrightarrow{\alpha }}. \label{eq:Jg} \eea
This expression reduces into the fermionic limit (Section IV) when
$M=2$. The $M=3$ case is exemplified in Appendix C.

The necessary condition (\ref{eq:condab}) is an imperative
step towards identifying normal transport (Fourier) systems, as it helps
us pinpoint current conserved systems directly, without detailed
numerical calculations.
If the system satisfies $\overrightarrow{\epsilon }\cdot \overrightarrow{\alpha }\neq \pm
\overrightarrow{\epsilon }\cdot \overrightarrow{\beta }$, one can
directly deduce that the thermal current is not conserved.
Note that in the Heisenberg model $F(\lambda)=0$, and only the next term in  Eq. (\ref{eq:Flambda})
is finite, accounting for dissipation of energy \cite{Vd}.
% The condition (\ref{eq:cond}) contains
%two parts: a requirement for relation between roots $\overrightarrow{\alpha }%
%\ $and $\overrightarrow{\beta }$, and a requirement for the local
%spectrum structure. (XXX)

%------------------------------------------------------------------
\section{Formal Fourier's law}

Recently, there are several ideas of how to approach Fourier's law
from fundamental principles \cite{Michel05,Michel06,
Pereira06,Antti,WuSegal07}. Here we will show that the appropriately
defined flux operator naturally leads to the discrete form of the law. The
derivation yields the conductivity coefficient for a general 1D
system in terms of the Hamiltonian parameters. We begin with a
generic nearest-neighbor Hamiltonian
\bea
H=\sum_{s} \left( h_s^0 +V(s,s+1) \right),
\eea
including local interactions and inter-site couplings.
In our definition (\ref{eq:j2}), the average flux,
$\overline j={\rm Tr}\{\rho j\}$, at weak interactions reads
\bea
\overline{j}_{s\rightarrow s+1}^{(2)} &=&
-\frac{ia}{2}{\rm Tr} \left\{ \rho [ (h_{s+1}^0-h_s^0), V(s,s+1)] \right\}
\nonumber\\
&=&-\frac{a}{2}{\rm Tr}\{\Delta h_{s}^{0}\Gamma
(t)\},
\eea
using the cyclic property of the trace.
Here $\rho$ is the total density matrix, $\Delta h_s^0=h_{s+1}^0-h_s^0$ is the difference
between local energies at neighboring sites and $ \Gamma (t)=i[V(s,s+1),\rho (t)]$ is hermitian.
We can also write this expression explicitly in terms of local $s$ functions,
\bea
\overline{j}_{s\rightarrow s+1}^{(2)}=-\frac{a}{2}({\overline{g}_{s+1}-\overline{g}_{s}}),
\eea
where
$\overline{g}_{s}={\rm Tr}\{h_{s}^{0}\Gamma (t)\}$.
%$\Gamma (t)=i[e_{s},\rho (t)]$ is hermitian.
If we define a local temperature $T_{s}$ at each site, we can then
relate the current between sites with the temperature difference $\Delta
T_{s}=T_{s+1}-T_{s}$,
\bea
\overline{j}^{(2)}_{s\rightarrow s+1}=-\left( a^2\frac{\Delta \overline{g}_{s}}{
{\Delta} \overline {h}^0_{s}}C_{s}\right)\frac {\Delta T_{s}}{a}
\label{eq:fourier} \eea
where
 $\Delta \overline{g}_{s}=\overline{g}_{s+1}-\overline{g}_{s}$,
and
$C_{s}=\frac{ {\Delta} {\overline h}^0_{s}}{\Delta T_{s}}$ is the
specific heat. This is the discrete Fourier's law
\cite{Michel06,Michel08}. We can identify the microscopic-local thermal
conductivity as $\kappa_s =a^2\frac{\Delta \overline{g}_{s}}{\Delta
\overline{h}_{s}^0}C_{s}$,
as long as $\overline{g}_{s}$ can be uniquely  defined (see discussion below), and  the ratio
$\frac{\Delta \overline{g}_{s}}{\Delta \overline{h}^0_{s}}$ is finite.
% page 20 in Michel06

As an example we consider a three-spin system. For the \textit{XY} model, if
the initial state is $\left\vert 0\right\rangle _{1}\left\vert
1\right\rangle _{2}\left\vert 0\right\rangle _{3}$,
%which is invariant under the exchange $P_{1,3}$,
it is easy to show that $\frac{\Delta \overline{g}
_{2}}{\Delta \overline{h}^0_{2}}=2\sqrt{2}\lambda \frac{\sin 2\sqrt{2}\lambda t
}{3\cos 2\sqrt{2}\lambda t+1}.$ For weak coupling, $\lambda t<1$, $\frac{
\Delta \overline{g}_{2}}{\Delta \overline{h}^0_{2}}\rightarrow 2\lambda ^{2}t$
holds.
%for an arbitrary initial diagonal density matrix (XXX).
The heat conductivity is then given by $\kappa =2\lambda
^{2}tC_{s}$, in agreement with our recent calculation
\cite{WuSegal07}. It also shows that although the total current of
the \textit{XY} model is conserved, the partial current between two
sites may have the form of the Fourier's law before thermal
equilibrium sets \cite{Bonetto00}.

We explain next how to define $\overline g_s$ uniquely. Although we
could formally write Eq. (\ref{eq:fourier}) , $\overline{g} _{s}$
may not be uniquely defined because $\Gamma (t)$ depends on the
index $s$: $\Gamma (t)$ could be either defined as $i[V(s,s+1),\rho
(t)]$ or $i[V(s,s-1),\rho (t)]$. Therefore, the condition for
$\overline{g}_{s}$ to be exclusively defined is
\bea
\mathrm{Tr}_{s}\{[V(s,s+1),\rho (t)]\}=\mathrm{Tr}_{s}\{[V(s,s-1),\rho (t)]\}.
\label{eq:Trs}
\eea
The trace $\mathrm{Tr}_{s}$ runs over all sites except site $s$. It is easy
to show that $P_{s-1,s+1}\rho (t)P_{s-1,s+1}=\rho (t)$ is a sufficient condition
for satisfying Eq. (\ref{eq:Trs}),
where $P_{s-1,s+1}$ is the exchange operator between sites $s-1$ and $s+1$.
If the total Hamiltonian is invariant under $P_{s-1,s+1}$, as it is in many
physical cases, the last condition translates into a condition on the system preparation,
\bea
P_{s-1,s+1}\rho (0)P_{s-1,s+1}=\rho(0).
\label{eq:Ex}
\eea
This is a sufficient (but not necessary) condition for attaining a unique
expression for $\overline g_s$.
Once $\overline g_s$ is carefully defined, we can proceed and calculate the thermal conductivity
using Eq. (\ref{eq:fourier}).
In the example above the initial state was set to
$\left\vert 0\right\rangle _{1}\left\vert
1\right\rangle _{2}\left\vert 0\right\rangle _{3}$,
which is indeed invariant under the exchange $P_{1,3}$.
Note that since the validity of the Fourier's law is independent of initial conditions,
the requirement to fulfill Eq. (\ref{eq:Ex}) is solely meant for distinctively identifying
the conductivity.

%--------------------------------------------------------------
\section{Conclusion}

In this paper we present and re-examine the heat flux operator that exactly satisfies
the continuity equation for  a general Hamiltonian in one
dimension. Based on the definition, we  deduce the necessary
conditions on the inter-site interaction that result in current
conservation.
This analysis  sets the first step towards the exploration of the
validity of Fourier's law of heat conduction in Hamiltonian systems:
systems that conserve energy have diverging conductivity.
As an example, using a simple operator algebra,
we  prove that independent bosons and fermions conduct heat ballistically.
We further apply the definition to various impurity
models, relevant for understanding heat flow in nanojunctions, and
obtain a non-perturbative non-stationary expression for the heat
current. The microscopic heat conductivity coefficient naturally emerges in the
present definition.

While previous works have typically relied on specific
quantum states, calculating only expectation values, see for example
\cite{Saito03,Casati05,Gemmer06,Gemmer07,Michel08}, the results
presented here essentially depend only on operator calculations.
Possible extensions include generalization of the heat current definition to time
dependent situations, and exploration of the necessary condition for the
applicability of the Fourier's law of heat conduction in 1D chains
\cite{WuSegal07}.

\begin{acknowledgments}
This project was supported by NSERC and by the University of Toronto Start-up Fund.
\end{acknowledgments}

%--------------------------------------------------------------------

\renewcommand{\theequation}{A\arabic{equation}}
\setcounter{equation}{0}  % reset counter
\section*{APPENDIX A: An Alternative, symmetry limited, definition for the heat current }
% use *-form to suppress numbering

We follow here a symmetry-limited definition of the heat flux operator
often adopted in the literature \cite{Michel06,Gemmer06, Gemmer07}.
The generic 1D Hamiltonian includes local potentials and inter-site
interactions
\bea
H=\sum_s{ (h_s^0} + V(s,s+1)).
\eea
The heat flux operator is defined by considering the time evolution
of the {\it local, non-interacting} energy operator,
\bea
\frac{dh_s^0}{dt}&=&i[H,h_s^0]
\nonumber\\
&=& -i[h_s^0,V(s,s-1)] -i[h_s^0,V(s,s+1)].
\label{eq:Hs}
\eea
We next assume that a continuity equation for $h_s^0$ holds,
based on the approximation that the local energy is conserved \cite{Gemmer06}
\bea
\frac{dh_s^0}{dt}= \frac{(j_{s-1}-j_s)}{a}.
\label{eq:J}
\eea
By comparing Eq. (\ref{eq:Hs}) with (\ref{eq:J}) one can identify the
current between sites as
\bea
j_{s}=ia[h_s^0,V(s,s+1)]; \,\,\,\ j_{s-1}=-ia[h_s^0,V(s,s-1)].
\nonumber\\
\label{eq:defl}
\eea
However, the second equality above produces
$j_s=-ia[h_{s+1}^0,V(s+1,s)]$ when shifted to site $s$. This can
be consistent with the first equality of Eq. (\ref{eq:defl})
only if the condition
\bea
[h_s^0+h_{s+1}^0,V(s,s+1)]=0
\label{eq:cond}
\eea
is satisfied. The definition (\ref{eq:defl}) is thus restricted to a
limited class of Hamiltonians that satisfy (\ref{eq:cond}). We
emphasize again that the heat current was defined here by studying
{\it local, non-interacting} energy changes, while Eq. (\ref{eq:js})
defines the heat current by studying the total energy at a site,
incorporating inter-site interactions, see Eq. (\ref{eq:eq4}). The
Heisenberg spin-$\frac{1}{2}$ model, $h_s^0=
\frac{\epsilon}{2}\sigma_s^z$, $V(s,s+1)= \lambda(\sigma_s^x
\sigma_{s+1}^x + \sigma_s^y \sigma_{s+1}^y + \sigma_s^z
\sigma_{s+1}^z)$, is an example of a system obeying (\ref{eq:cond}).

%----------------------------------------------------------------------

\renewcommand{\theequation}{B\arabic{equation}}
\setcounter{equation}{0}  % reset counter
\section*{APPENDIX B: Spin-Boson model:
Derivation of the weak coupling expression for the heat current}

We derive here a weak-coupling expression for the steady-state
heat current in the spin-boson model
using the non-perturbative definition (\ref{eq:JSB}).
The two-bath ($\nu=L,R$) spin-boson Hamiltonian is given by
\bea H_{SB}=H_{spin}+\sum_{\nu}H^0_{\nu} + \sum_{\nu}V_{\nu},
\eea
where
\bea H_{spin}= \frac{B}{2}\sigma^z;\,\, H^0_{\nu}=\sum_{q}\omega_q
b_{\nu,q}^{\dagger} b_{\nu,q}; \,\,
V_{\nu}= \sigma^x X_{\nu}.
\eea
Here  $B$ is the spin splitting, $b_{\nu,q}^{\dagger}$ is a creation
operator satisfying the bosonic statistics, and $V_{\nu}$ includes
system-bath interactions at each contact, $X_{\nu}=\sum_{q}
\lambda_{\nu,q}(b_{\nu,q}^{\dagger}+b_{\nu,q})$. There is no direct
coupling between the two harmonic baths (temperature $T_{\nu}$), as
they are coupled only through the central spin.

The general expression for the current operator is given by Eq.
(\ref{eq:j2}), $j_{L}=\frac{i}{2}[H^0_{L}-H_{spin},V_L]$,
disregarding for convenience the factor $a$. Note that $j^{(4)}=0$,
see Eq. (\ref{eq:j4}), since $[V_L,V_R]=0$. In the present model
the current operator from the $L$ interface to the spin is given by
\bea j_{L}=\frac{1}{2}[B\sigma^y X_L + \sigma^x P_L],
\label{eq:C0} \eea
where $P_L=i\sum_{q}\lambda_{L,q}\omega_q(b_{L,q}^{\dagger}-b_{L,q})$
denotes the  sum of the momenta of the harmonic oscillators at the
left boundary. This expression is valid in the non-perturbative
regime and for non-stationary situations.
In steady-state the expectation value of the interaction is zero, e.g. at the $L$ contact,
\bea \Big\langle \frac{\partial V_L}{\partial t}\Big\rangle =
\langle \dot\sigma^x X_L+\sigma^x \dot X_L\rangle=0. \eea
Since $\dot \sigma^x=-B \sigma^y$ and  $\dot X_L=P_L$, we find that
$\langle \sigma^xP_L\rangle=\langle B\sigma^yX_L \rangle$. The
stationary heat current is therefore given by
\bea
\langle j_ {L}\rangle= {\rm Tr}\{\rho
j_{L}\}= B {\rm Tr}\{\rho \sigma^yX_L\},
\eea
where $\rho$ is the total density matrix.
Using the energy representation, $\sigma^z=|u\rangle \langle
u|-|d\rangle \langle d|$, $\sigma^x=|d\rangle \langle u|+|u\rangle
\langle d|$, $\sigma^y=-i|u\rangle \langle d|+i|d\rangle \langle
u|$, we can write the heat current as
\bea
\langle j_ {L}\rangle= iB {\rm
Tr_B}\{(\rho_{u,d}-\rho_{d,u})X_L\},
\label{eq:C1}
\eea
where
 ${\rm Tr_B}$ denotes the trace over the thermal baths ($L$ and $R$) states only.
 This expression can be evaluated by solving the
Liouville equation, written here explicitly for the nondiagonal matrix
element
\bea
%&&\dot\rho_{1,1}=-iX\rho_{0,1} +i\rho_{1,0}X
%\nonumber\\
%&&\dot\rho_{0,0}=-iX\rho_{1,0} +i\rho_{0,1}X
%\nonumber\\
%&&
\dot\rho_{d,u}(t)=iB\rho_{d,u}(t) -iX(t)\rho_{u,u}(t) +i \rho_{d,d}(t)X(t), \eea
with $X=X_L+X_R$.
Formal integration of this differential equation yields
\bea
\rho_{d,u}(t)=
\int_0^{t} e^{iB(t-\tau)} [-iX(\tau)\rho_{u,u}(\tau) +i\rho_{d,d}(\tau)X(\tau)]d\tau.
\nonumber\\
\label{eq:rhodu}
\eea
We evaluate next the term ${\rm Tr_B}\{\rho_{d,u}X_L\}$ under the
following approximations: (i) weak system-bath coupling, neglecting
higher order correlation functions, (ii) Markovian limit, assuming
the spin's relaxation timescale is longer than that of the bath
fluctuations, and (iii) initial factorized condition, where $\rho$
is well approximated by the product
$\rho(t=0)=\rho_{spin}(t=0)\rho_L \rho_R$. Here
$\rho_{\nu}=e^{-H_{\nu}^0/T_{\nu}}/{\rm Tr}
\{e^{-H_{\nu}^0/T_{\nu}}\}$ are the density operators of the thermal
baths. These assumptions are compatible with the Redfield
approximation \cite{Redfield}. Using (\ref{eq:rhodu}) we get
 \bea {\rm
Tr_B} \{\rho_{d,u}X_L\}&=&-ip_u(t) \int_0^{\infty} e^{iB\tau}\langle
X_L(\tau)X_L(0)\rangle d\tau
\nonumber\\
&+&ip_d(t)\int_0^{\infty} e^{iB\tau}\langle X_L(0)X_L(\tau)\rangle
d\tau,
\label{eq:n1}
\nonumber\\ \eea
where $p_{u}={\rm Tr_B}\{\rho_{u,u}\}$ denotes  the population of the
spin-up state and $p_d$ is the spin-down population.
Note that terms of the form $\langle X_L(t) X_R(\tau)\rangle$ disappear,
since the two reservoirs are not correlated.
Following the same procedure for the second term in Eq. (\ref{eq:C1}) we obtain
\bea {\rm Tr_B} \{\rho_{u,d}X_L\}&=&ip_u(t) \int_{-\infty}^{0}
e^{iB\tau}\langle X_L(\tau)X_L(0)\rangle d\tau
\nonumber\\
&-&ip_d(t)\int_{-\infty}^{0} e^{iB\tau}\langle
X_L(0)X_L(\tau)\rangle d\tau. \nonumber\\
\label{eq:n2}
\eea
Combining equations (\ref{eq:n1}) and (\ref{eq:n2}) provides us with
the stationary thermal current under weak-coupling and Markovian
approximations,
\bea
\langle
j_{L}\rangle= -B[p_u k_{u\rightarrow d}^L - p_d
k_{d\rightarrow u}^L],
\label{eq:C2}
\eea
with the relaxation rates
\bea &&k_{u\rightarrow d}^L=\int_{-\infty}^{\infty} e^{iB\tau}
\langle
X_L(\tau)X_L(0)\rangle d\tau \nonumber\\
&&
k_{d\rightarrow u}^{\nu}=\int_{-\infty} ^{\infty}e^{-iB\tau}\langle
X_{\nu}(\tau)X_{\nu}(0)\rangle d\tau.
%k_{d\rightarrow u}^L=k_{u\rightarrow d}^L e^{-B/T_L}.
\label{eq:C3}
\eea
Equation (\ref{eq:C2}) describes energy current through the
junction, calculated e.g. at the $L$ contact, as the balance between
an energy gain from the reservoir to the spin, and an energy loss
from the spin to the $L$ bath.

The diagonal elements of the density matrix, $p_d$ and
$p_u$, can be further calculated under the same set of approximations, to yield  the
quantum Master equation,
\bea
&& \dot p_u=-p_u(t) \sum_{\nu}k_{u\rightarrow d}^{\nu}  + p_d(t) \sum_{\nu} k_{d\rightarrow u}^{\nu}
\nonumber\\
&&p_u(t)+p_d(t)=1.
\eea
In steady state ($\dot p =0$) the spin occupations are
\bea
&&p_u= \frac{k_{d\rightarrow u}^L+ k_{d\rightarrow
u}^R}{k_{d\rightarrow u}^L +  k_{d\rightarrow u}^R + k_{u\rightarrow
d}^R +k_{u\rightarrow d}^R }; \nonumber\\
&&p_u+p_d=1.
\label{eq:popul}
\eea
Plugging Eq. (\ref{eq:popul}) into (\ref{eq:C3}) leads to an explicit expression for the current
\bea
\langle j_{L}\rangle  = \frac{k_{u\rightarrow d}^L
k_{u\rightarrow d}^R (e^{-B/T_L} - e^{-B/T_R})}{ k_{u\rightarrow
d}^R + k_{u\rightarrow d}^L + k_{d\rightarrow u}^R + k_{d\rightarrow
u}^L}.
\eea
An analogous expression holds at the $R$ contact.
This is the well established quantum Master-equation limit, used in
various applications \cite{SegalRectif,radiation,SegalNDR,SegalPump,Persson,Mukamel,Pekola}.

We can also extend the calculations to non-stationary situations.
In this case one needs to evaluate the extra term ${\rm
Tr}\{\sigma^xP_L\}={\rm Tr_B}\{(\rho_{u,d}+\rho_{d,u})P_L\}$ in Eq. (\ref{eq:C0}), resulting
in momentum-position correlation functions of the form,
 $\langle P_L(t) X_L(\tau)\rangle$ in second order system-bath couplings.

Note that $[H_L^0+H_{spin},V_L] \neq 0$ for the spin-boson Hamiltonian. Therefore,
we cannot use in general the definition of Appendix A,  $j_{L}=i[H_{spin},V_L]$.
This limited expression is still applicable in a steady-state situation
since $\langle \partial V_L /\partial t\rangle =0$,  translating into
$\langle [H_{SB},V_L] \rangle = \langle [H_L^0+H_{spin},V_L] \rangle=0$,
see Eq. (\ref{eq:cond}).

%------------------------------------------------------------------------

\renewcommand{\theequation}{C\arabic{equation}}
\setcounter{equation}{0}  % reset counter
\section*{APPENDIX C: Current conservation in an $M$=3 states model }
 We clarify the notation and the results of Section V using an
$M$=3 level system.  According to our notation, the diagonal
operators are
\bea \overrightarrow{h}=(\left|1\right\rangle \left\langle 1\right|
,\left| 2\right\rangle \left\langle 2\right|,  \left|
3\right\rangle\left\langle3\right|). \eea
The six nondiagonal operators with their respective roots are
\bea &&E^{\overrightarrow{ \alpha }_{1}}=\left| 1\right\rangle
\left\langle 2\right|, \,\,\, \overrightarrow{\alpha }_{1}=(1,-1,0)
 \nonumber\\
&&E^{\overrightarrow{\alpha }_{2}}=\left| 2\right\rangle
\left\langle 1\right| , \,\,\, \overrightarrow{\alpha }_{2}=(-1,1,0)
\nonumber\\
&&E^{\overrightarrow{ \alpha }_{3}}=\left| 1\right\rangle
\left\langle 3\right|,\,\,\, \overrightarrow{\alpha}_{3}=(1,0,-1)
\nonumber\\
&&E^{\overrightarrow{ \alpha }_{4}}=\left| 3\right\rangle
\left\langle 1\right|,\,\,\, \overrightarrow{\alpha}_{4}=(-1,0,1)
\nonumber\\
&&E^{\overrightarrow{ \alpha }_{5}}=\left| 2\right\rangle
\left\langle 3\right|,\,\,\, \overrightarrow{\alpha}_{5}=(0,1,-1)
\nonumber\\
&&E^{\overrightarrow{ \alpha }_{6}}=\left| 3\right\rangle
\left\langle 2\right|,\,\,\, \overrightarrow{\alpha}_{6}=(0,-1,1),
\eea
where the energies at each site are
$\overrightarrow\epsilon=(\epsilon_1,\epsilon_2,\epsilon_3)$. If the
system conserves current [i.e. it fulfills (\ref{eq:condab})],  the
site-site interaction can include only the following terms:
$E_s^{\overrightarrow\alpha_n} E_{s+1}^{\overrightarrow\alpha_n}$,
($n=1..6$), and the pairs $E_s^{\overrightarrow\alpha_1}
E_{s\pm1}^{\overrightarrow\alpha_2}$, $E_s^{\overrightarrow\alpha_3}
E_{s\pm1}^{\overrightarrow\alpha_4}$, $E_s^{\overrightarrow\alpha_5}
E_{s\pm1}^{\overrightarrow\alpha_6}$. The current operator in this
model is given by Eq. (\ref{eq:Jg}),
\bea &&j_{s\rightarrow s+1}^{(2)}=ia\bigg\{
V_{\overrightarrow\alpha_1,\overrightarrow\alpha_2}(\epsilon_1-\epsilon_2)(E_s^{\overrightarrow\alpha_1}E_{s+1}^{\overrightarrow\alpha_2}
-E_s^{\overrightarrow\alpha_2}E_{s+1}^{\overrightarrow\alpha_1})
\nonumber\\
&&+
V_{\overrightarrow\alpha_3,\overrightarrow\alpha_4}(\epsilon_1-\epsilon_3)(E_s^{\overrightarrow\alpha_3}E_{s+1}^{\overrightarrow\alpha_4}
-E_s^{\overrightarrow\alpha_4}E_{s+1}^{\overrightarrow\alpha_3})
\nonumber\\
&&+
V_{\overrightarrow\alpha_5,\overrightarrow\alpha_6}(\epsilon_2-\epsilon_3)(E_s^{\overrightarrow\alpha_5}E_{s+1}^{\overrightarrow\alpha_6}
-E_s^{\overrightarrow\alpha_6}E_{s+1}^{\overrightarrow\alpha_5})
\bigg\},
\nonumber\\
 \eea
a generalization of the spin chain result (\ref{eq:spinJ}).
%-------------------------------------------------------------------------


\begin{thebibliography}{99}
\bibitem{SB}
After applying the small-polaron transformation \cite{Mahan}, we obtain the flux operator
%
\bea
j_{L}=\frac{\Delta}{2}\big ( B\sigma^y e^{i\Omega_L}
+i \sigma ^x \sum_q \omega_q[b_{L,q}^{\dagger}b_{L,q}, e^{i\Omega_L} ]
\big),
\eea
%
where $\Omega_L=i\sum_q\frac{\kappa_{L,q}}{\omega_k} (b_{L,q}^{\dagger}-b_{L,q})$.
\bibitem{noteE}
For a two level system
$E^{ \overrightarrow{\alpha }}=\sigma^{\pm}$, thus $F(\lambda)=0$ when the inter-site interaction
includes the following terms $\sigma_s^+ \sigma_{s+1}^-$, $\sigma_s^+ \sigma_{s+1}^+$, and
$\sigma_s^- \sigma_{s+1}^-$.
\bibitem{Vd}
In first order of $\lambda$ the commutator $[j_s,H]=
[j^{(2)}_{s\rightarrow s+1},h_s^0+h_{s+1}^0 ]$ where
$j^{(2)}_{s\rightarrow s+1}\propto [h_s^0-h_{s+1}^0,V(s,s+1)]$.
Since $V_d \propto Un_s^i n_{s+1}^j$ commutes with $h_s^0\propto
n_s^i$, the many body $V_d$ term does not destroy current
conservation in the lowest approximation.
\end{thebibliography}
\end{document}